# On the Dynamics of the Sagittarius Dwarf Galaxy

Héctor Velázquez[1]* and Simon D.M. White[1]*
[1]*Institute of Astronomy, Madingley Road, Cambridge CB3 0HA*

astro-ph/9503022    5 Mar 1995

**ABSTRACT**
We use numerical simulations to test the feasibility of the suggestion by Ibata et al. (1994) that the excess population of stars which they discovered in the Sagittarius region may be the disrupted remains of a dwarf spheroidal galaxy. We find that a Fornax-like model for the pre-disruption system can indeed reproduce the data. However, the galaxy must be on a relatively short period orbit with a pericentre of about 10 kpc and an apocentre of about 52 kpc, giving a current transverse velocity of 255 km/s and a period of $\sim 760$ Myr. Furthermore, disruption must have occurred predominantly on the last pericentric passage rather than on the present one. The data are consistent with transverse motion either towards or away from the Galactic Plane. These results depend primarily on the rotation curve of the Galaxy and are insensitive to the mass distribution in its outer halo or to the mass of its disk.

**Key words:** galaxy: kinematics and dynamics – galaxy: individual Sagittarius dSph – methods: numerical

## 1 INTRODUCTION

Recently Ibata, Gilmore and Irwin (1994) discovered a population of stars in the Sagittarius region which appears to be a dwarf spheroidal (dSph) galaxy in process of total disruption. This population extends from $\ell = 10^o$ to $\ell = 5^o$ and from $b = -20^o$ to $b = -12^o$ (in galactic coordinates) and is elongated almost perpendicular to the Galactic Plane. From its size, luminosity and stellar content Ibata et al. suggest that the Sagittarius dSph galaxy was comparable to Fornax before its disruption.

These stars all have heliocentric radial velocities near +140 km/s and are located at a distance of about 24 kpc from the Sun. Furthermore, the observed mean radial velocity difference between stars at $b = -12^o$ and stars at $b = -20^o$ is less than 5 km/s (Ibata et al. 1994). This appears to be the nearest known dwarf galaxy, and it provides us with an opportunity to study in close-up how tidal processes can disrupt a satellite and incorporate it into our Galaxy.

The purpose of this paper is to examine the dynamics of Sagittarius population on the hypothesis that it does indeed correspond to a dSph galaxy. To this end we carry out numerical simulations of the disruption process and compare the results with the observational data. We conclude that the dSph model can describe the data well, but that this requires an initial orbit of surprisingly short period.

* Present address: Max-Planck-Institut für Astrophysik, Karl-Schwarzschild-Strasse 1, D-85740 Garching bei München, Germany.

**Table 1.** Parameters for the Milky Way models

| Model: | Symbol | Physical Units |
|---|---|---|
| Galactic Disk: | | |
| | $M_d$ | $5.5 \times 10^{10}$ M$_\odot$ |
| | h | 3.5 kpc |
| | $z_o$ | 350 pc |
| Galactic Bulge: | | |
| | $M_b$ | $1.1 \times 10^{10}$ M$_\odot$ |
| | a | 525 pc |
| Galactic Halo: | | |
| | $\gamma$ | 3.5 kpc |
| (Model 1) | $M_h$ | $5.3 \times 10^{11}$ M$_\odot$ |
| | $r_c$ | 77 kpc |
| (Model 2) | $M_h$ | $1.5 \times 10^{12}$ M$_\odot$ |
| | $r_c$ | 200 kpc |

## 2   THE NUMERICAL SIMULATIONS

We assume a rigid model for the potential of our Galaxy. It consists of three components. The disk is given by the density profile (Quinn & Goodman 1986)

$$\rho_d(R, z) = \frac{M_d}{4\pi h^2 z_o} \exp(-R/h - |z|/z_o) \quad (1)$$

where $M_d$ is its total mass, and $h$ and $z_o$ are the radial and vertical scale lengths, respectively. The bulge and halo are described by the following spherical models (discussed in detail by Hernquist 1990, Hernquist 1993)

$$\rho_b(r) = \frac{M_b}{2\pi} \frac{a}{r(r+a)^3} \quad (2)$$

and

$$\rho_h(r) = \frac{M_h \alpha}{2\pi^{3/2}} \frac{\exp(-r^2/r_c^2)}{r^2 + \gamma^2} \quad (3)$$

where $M_b$ and $a$ are the mass and scale-length of the bulge, $M_h$, $r_c$ and $\gamma$ are the mass, cutoff radius and core radius of the halo, respectively, and $\alpha$ is a normalization constant. We begin by studying two major sets of parameters which are summarized in Table 1. Model 1 has a minimal halo and has an escape velocity from the Solar neighbourhood of about 500 km/s (Carney & Latham 1987). Model 2 has a massive halo of the kind suggested by studies of the distribution of satellites around the Milky Way (Zaritsky et al. 1989) and around other giant spiral galaxies (Zaritsky & White 1994). Both models give a total circular rotation velocity of 220 km/s at the Solar radius.

We represent the pre-disruption dwarf by a King model (King 1966) with size parameters chosen to match those of the Fornax system. Specifically we choose core and tidal radii of 527 pc and 2735 pc, respectively, (Irwin & Hatzidimitriou 1993). We use the mass of $1.2 \times 10^8$ M$_\odot$ inferred for Fornax by Mateo et al. (1993) as a reference value for our models, but test a variety of possible masses for the pre-disruption system.

We tabulate the potential for our Galaxy models on a two-dimensional mesh $(R, z)$ with sufficient resolution to compute the forces on stars within the dSph by simple bilinear interpolation. Forces between stars in the dSph are computed using an N-body tree code with a tolerance parameter of $\theta = 0.8$ and with the quadrupole terms included (Hernquist 1987). The dSph galaxy is represented by 8192 equal mass particles, and we use a softening parameter of $\epsilon = 105$ pc. The dSph is allowed to relax in isolation before it is placed on an orbit within one of our Galactic potential models.

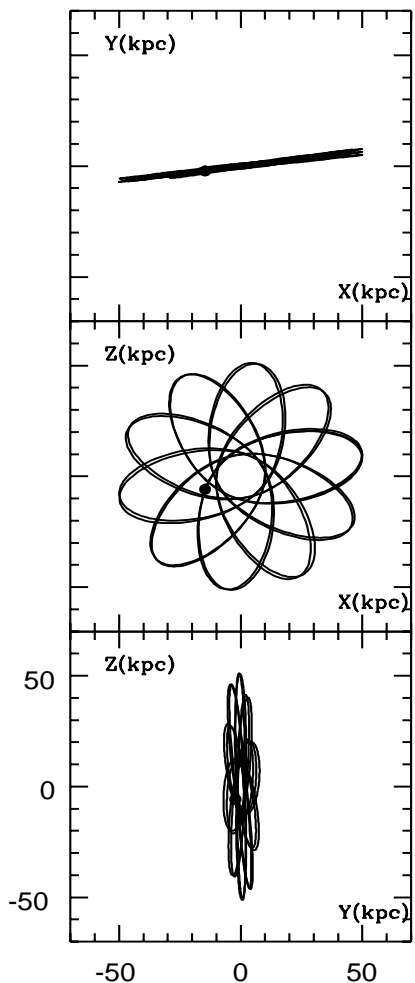

**Figure 1.** Orbit of the Sagittarius dwarf for Model 1 of our Galaxy. The Galaxy is centred at the coordinate origin with its disk in the $x$-$y$ plane. The north Galactic pole points towards positive $z$. The Sun is located at $(8.5, 0, 0)$ kpc. The orbit is plotted for a total time interval of 15 Gyr ending at its present position (filled circle).

The strong elongation of the Sagittarius population perpendicular to the Galactic plane suggests that its transverse motion must lie in this direction. Since the radial velocity and the distance of the population are known, the only parameter needed to specify its orbit completely is the magnitude (and sign) of the transverse motion. We determine this quantity using the fact that there is no detectable difference between the mean velocity of stars at $b = -12°$ and that of stars at $b = -20°$. The stars from a disrupting dwarf are expected to follow orbits very close to that of the barycentre of the system. We therefore examine test particle orbits which pass through the observed position with the observed radial velocity but have different transverse velocities. We then find the value of the transverse velocity for which the orbit shows no perceptible heliocentric radial velocity gradient between $b = -12^o$ and $b = -20^o$. Note that this method cannot predict the *sign* of the transverse velocity but that it predicts the magnitude quite accurately. Once the orbit is determined, we integrate it back in time for 1 Gyr and use the resulting position and velocity as initial conditions for a forward integration with an N-body satellite. We can then compare the present-day distribution of stars in the simulation with the observational data and verify *post hoc* that our choice of orbital parameters was appropriate.

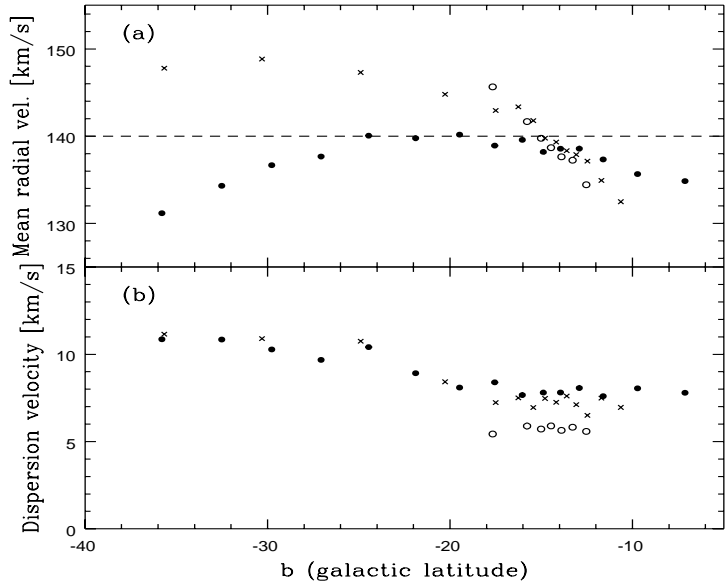

**Figure 2.** (a) Present mean heliocentric radial velocity is plotted against Galactic latitude for a simulation in which the Sagittarius dwarf initially had a mass of $9.8 \times 10^7$ $M_\odot$ and orbits within our model 1 for the Galaxy. Filled circles give the results found when the velocities of all satellite stars in this region are averaged, while unfilled circles correspond to using only stars which are still bound to the satellite. Crosses give the mean velocities for a more elongated orbit with a tangential velocity of $-285$ km/s (see text for more details). The dashed line is the observed mean velocity, $+140$ km/s. (b) Velocity dispersion as a function of Galactic latitude for the same cases shown in (a) In both cases, the data were binned according to Galactic latitude, and each bin has either 196 particles (filled circles and crosses) or 92 particles (unfilled circles).

Notice that our choice of a rigid model for the Galaxy means that dynamical friction is ignored in our simulations. For a $\sim 10^8 M_\odot$ satellite in a $\sim 1$ Gyr orbit, the effects of dynamical friction is expected to be at the one percent level (Binney & Tremaine 1987). Furthermore, we are, in practice, only interested in the last 1 Gyr or so during which the satellite is disrupted.

## 3   RESULTS AND DISCUSSION

Figure 1 shows the orbit we predict for the Sagittarius dwarf for Model 1 of the Galaxy. It is characterized by pericentric and apocentric distances of 10 kpc and 52 kpc, respectively, and it has an orbital period of about 760 Myrs. This orbit was obtained for a test particle with a current tangential velocity of $-255$ km/s (i.e. directed away from the Galactic plane).

In figures 2 and 3 we present the results of a forward simulation for a satellite of initial mass $9.8 \times 10^7$ M$_\odot$ placed on this orbit. The evolution was followed for the last 1 Gyr and so included two pericentric passages. Figure 2(a) shows that at the present day the population of satellite stars shows no perceptible trend of the mean heliocentric radial velocity with Galactic latitude in the interval between $b = -20^o$ and $b = -12^o$ (filled circles). However, if only the stars which are still bound to the satellite are considered then a clear velocity gradient is found (unfilled circles). This difference may be explained by noting that stripped material on the leading arm is pulled toward the Galactic centre and to higher velocity, while trailing material is pulled away (Oh, Lin and Aarseth 1994). This cancels the apparent velocity gradient across the bound system due to its finite angular extent. Such cancellation only occurs for orbits close to the one shown in Figure 1. For example we show results in Figure 2 for a slightly more elongated orbit corresponding to a present tangential velocity of $-285$ km/s, giving pericentric and apocentric distances of 10 kpc and 68 kpc, respectively, and an orbital period of about 1 Gyr. This orbit leads to a velocity gradient which is $> 1$ km/s/deg. Thus, a change of $\pm 30$ km/s in the tangential velocity already produces a stronger gradient than is allowed by the data. Figure 2(b) shows that our best model gives a velocity dispersion of about 8 km/s over the whole Galactic latitude range $b = -12^o$ to $b = -20^o$. This is somewhat smaller than the observed value of 10 km/s, suggesting that we should increase somewhat the mass of our satellite. In our model the mean radial velocity of stars at $b = -30^o$ is predicted to be $+136$ km/s. This population should be easily detectable and so should provide a test of the model. Contours of the star counts (fig. 3) show that the satellite in our model has only a weak central concentration and is strongly elongated perpendicular to the Galactic plane. Thus both the spatial and the kinematic structure are in good agreement with the data for the Sagittarius population presented by Ibata et al. (1994).

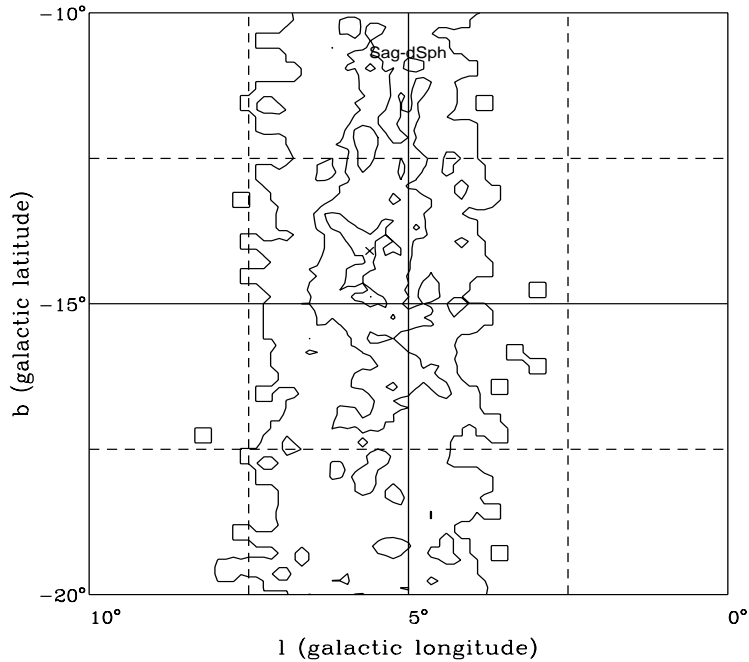

**Figure 3.** Isodensity contours of star counts for the simulation of figs 1 and 2. The three contours corrrespond to surface densities which are 9/13, 5/13 and 1/13 of the maximum value which occurs at the point labelled ×. This figure is very similar to the one shown by Ibata et al. 1994.

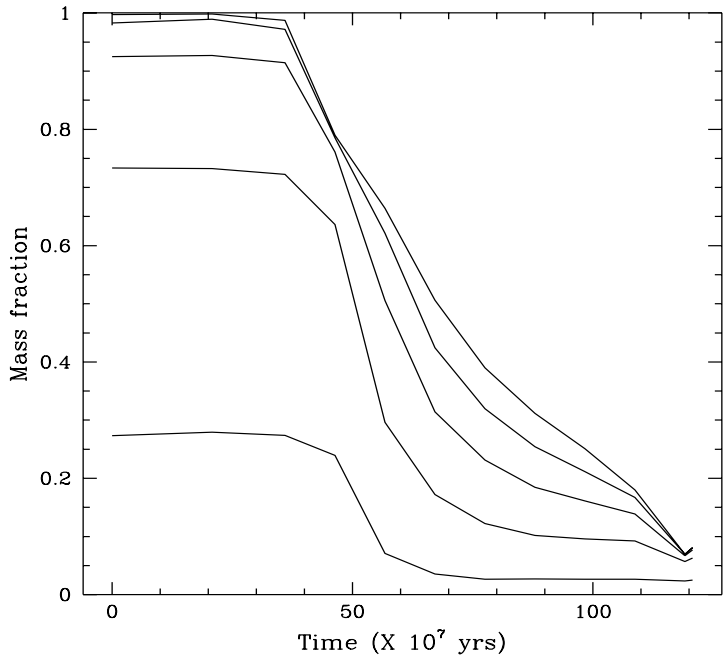

**Figure 4.** The evolution of the mass fraction within spheres of various radii for the model of figs 1 to 3. The spheres are centred on the unperturbed orbit and have radii of 0.527, 1.054, 1.581, 2.108, 2.635 kpc (from bottom to top). The ordinate is the ratio of the contained mass to the initial total mass of the satellite.

The strong elongation seen in fig. 3 is a result of the tidal distortion of the dwarf. Indeed it is almost entirely disrupted at this time. We illustrate this in fig. 4 which shows how the mass contained within a variety of radii varies with time. It is clear from this plot that the main damage to the satellite occurs during pericentric passages. After the first one the inner regions appear to approach a new equilibrium but the outer parts continue to expand as tidal effects string stars out along the orbit. This process accelerates as the dwarf approaches pericentre for the second time. After this second encounter only about 8% of the initial mass remains bound to the satellite.

During this 1 Gyr the satellite crosses the disk three times. Fig. 4 shows no clear indication of these events, suggesting that the evolution of the satellite is not sensitive to the internal structure of the disk. To check this, we ran a simulation which replaced the disk by an additional spherical component which gave the same rotation curve. The results were almost indistinguishable. This may be because the most central crossing of the disk occurs at a galactocentric distance of about 3.5 radial scale lengths of the disk. At such distances the disk potential is only moderately flattened. Thus, the evolution of the satellite is determined almost entirely by the global potential of the Galaxy, and so by its rotation curve.

We ran similar simulations using Model 2 for the Galaxy halo. However, we found that this made very little difference to the results. Slight decreases (50 kpc and 750 Myrs) in the apocenter and period of the orbit were needed, and the present bound mass rose to about 11% of the initial mass. We also tried orbits in both model potentials in which the present transverse motion was reversed. Again the results obtained were very similar to those shown in figs 1 to 4. It is not possible to determine the sense of the tranverse motion from the presently available data.

Our results rule out the suggestion of Koribalski, Johnston and Robina (1994) that the Sagittarius dwarf galaxy is currently experiencing its first pericentric passage. Such models predict a heliocentric radial velocity difference of at least 30 km/s between $b = -12^o$ and $b = -20^o$ and are inconsistent with the observational data. Furthermore, the observed elongation of the density contours can only be obtained for implausibly low initial masses ($\leq 10^6 M_\odot$) on such an orbit.

Note that the dSph models which do fit the data of Ibata et al. (1994) require the satellite galaxy to survive at least ten pericentric passages before being almost entirely disrupted on the passage which immediately preceeded the present one. While at first sight this seems surprising, it must clearly be possible. Any satellite must suffer final catastrophic disruption on some pericentric passage, and in the case of Sagittarius we are seeing a system where this event occurred only after a series of previous less damaging encounters had reduced its mass and binding energy to the point of critical stability.

## 4   CONLUSIONS

Our main conclusions are:

(i) a Fornax-like galaxy is a feasible predecessor for the population of stars discovered by (Ibata et al. 1994); (ii) the orbit of this Sagittarius dwarf has pericenter and apocenter at galactocentric radii of about 10 kpc and 52 kpc, respectively; (iii) this orbit has a period of about 760 Myr so that Sagittarius completed more than 10 orbits before it was finally disrupted, (iv) the major disruption event occurred during the last pericentric passage rather than during the present one, (v) the present transverse velocity of Sagittarius is about 255 km/s but it is impossible to establish the direction of this transverse motion from the presently available data.

## ACKNOWLEDGMENTS

We wish to thank L. Hernquist for the use of his tree algorithm. We are also grateful to R. Ibata, M. Irwin and G. Gilmore for provide us with their data and to the referee for useful comments. Finally, HV wishes to thank CONACyT (Consejo Nacional de Ciencia y Tecnología) of México for financial support, and R. Ibata for helpful comments.